\documentclass[twocolumn,pra,aps,showpacs,superscriptaddress,floatfix,showkeys,nofootinbib,10pt]{revtex4-2}

\usepackage{graphicx,amsmath,amssymb,color}
\usepackage{float}
\usepackage{dcolumn}
\usepackage{bm}
\usepackage{physics}
\usepackage[table,xcdraw]{xcolor}
\usepackage{natbib}
\usepackage{amsthm}
\usepackage{comment}
\usepackage{booktabs}
\usepackage{braket}
\usepackage{amsmath}
\usepackage{mathtools}
\usepackage{dsfont}
\usepackage[colorlinks=true,linkcolor=blue,citecolor=magenta,urlcolor=blue]{hyperref}
\usepackage{soul}

\newtheorem{remark}{Remark}
\newtheorem{result}{Result}

\newcommand{\be}{\begin{equation}}
\newcommand{\ee}{\end{equation}}
\newcommand{\beq}{\begin{eqnarray}}
\newcommand{\eeq}{\end{eqnarray}}

\renewcommand{\va}{\mathbf{a}}
\renewcommand{\vb}{\mathbf{b}}
\newcommand{\vx}{\mathbf{x}}
\newcommand{\vy}{\mathbf{y}}

\renewcommand{\Pr}{p}
\renewcommand{\tr}{\mathrm{tr}}

\newcommand{\vi}{\mathrm{v}}

\newcommand{\veps}{\varepsilon}

\begin{document}


\title{Exponentially Decreasing Critical Detection Efficiency for Any Bell Inequality}


\author{Nikolai Miklin} 
\email{miklin@hhu.de}
\affiliation{Institute of Theoretical Physics and Astrophysics, National Quantum Information Center, Faculty of Mathematics, Physics and Informatics, University of Gdansk, 80-952 Gda\'nsk, Poland}
\affiliation{Heinrich Heine University D{\"u}sseldorf, Universit{\"a}tsstra{\ss}e 1, 40225 D{\"u}sseldorf, Germany}

\author{Anubhav Chaturvedi}
\email{anubhav.chaturvedi@phdstud.ug.edu.pl}
\affiliation{Institute of Theoretical Physics and Astrophysics, National Quantum Information Center, Faculty of Mathematics, Physics and Informatics, University of Gdansk, 80-952 Gda\'nsk, Poland}

\author{Mohamed~Bourennane} 
\email{boure@fysik.su.se}
\affiliation{Department of Physics, Stockholm University, S-10691 Stockholm, Sweden}

\author{Marcin Paw{\l}owski} 
\email{marcin.pawlowski@ug.edu.pl}
\affiliation{Institute of Theoretical Physics and Astrophysics, National Quantum Information Center, Faculty of Mathematics, Physics and Informatics, University of Gdansk, 80-952 Gda\'nsk, Poland}
\affiliation{International Centre for Theory of Quantum Technologies (ICTQT), University of Gdansk, 80-308 Gda\'nsk, Poland}

\author{Ad\'{a}n Cabello} 
\email{adan@us.es}
\affiliation{Departamento de F\'{i}sica Aplicada II, Universidad de Sevilla, E-41012 Sevilla, Spain}
\affiliation{Instituto Carlos I de F\'{i}sica Te\'{o}rica y Computacional, Universidad de Sevilla, E-41012 Sevilla, Spain}


\begin{abstract}
We address the problem of closing the detection efficiency loophole in Bell experiments, which is crucial for real-world applications. Every Bell inequality has a critical detection efficiency $\eta$ that must be surpassed to avoid the detection loophole. Here, we propose a general method for reducing the critical detection efficiency of any Bell inequality to arbitrary low values.
This is accomplished by entangling two particles in $N$ orthogonal subspaces (e.g., $N$ degrees of freedom) and conducting $N$ Bell tests in parallel.
Furthermore, the proposed method is based on the introduction of penalized $N$-product (PNP) Bell inequalities, for which the so-called simultaneous measurement loophole is closed, and the maximum value for local hidden-variable theories is simply the $N$th power of the one of the Bell inequality initially considered. We show that, for the PNP Bell inequalities, the critical detection efficiency decays {\em exponentially} with $N$. The strength of our method is illustrated with a detailed study of the PNP Bell inequalities resulting from the Clauser-Horne-Shimony-Holt inequality.
\end{abstract}


\maketitle


{\em Introduction.---}Quantum correlations arising from local measurements on entangled particles~\cite{Bell1964} allow for multiple applications, including device-independent randomness expansion~\cite{colbeck2009quantum,pironio2010random,liu2021device,shalm2021device}, quantum key distribution~\cite{Ekert1991,mayers1998quantum,BHK05,Acin2007,PABGMS09}, secret sharing~\cite{AGCA12,MBNC19}, self-testing~\cite{mayers2003self,supic2020selftestingof}, and certification of quantum measurements~\cite{GGGCBDXCKVL16,SMNPCB19,QBWCC19}. All these tasks require a loophole-free Bell test~\cite{Hensen2015,giustina2015significant,Shalm2015,rosenfeld2017event} as a necessary condition. The most challenging problem from the applications' perspective is closing the detection loophole \cite{Pearle1970}, since otherwise an adversary can simulate the behavior of entangled particles provided that a sufficient fraction of them remains undetected. Therefore, a fundamental problem is to identify quantum correlations that cannot be simulated with local hidden-variable (LHV) models even when the detection efficiency is relatively low.

The detection efficiency in a Bell inequality test is the ratio between the number of systems detected by the measuring devices and the number of systems emitted by the source. It depends not only on the properties of the detectors, but also on the losses in the channel.
Closing the detection loophole requires surpassing a certain threshold detection efficiency, which depends on the quantum correlations chosen. For symmetric Bell tests (i.e., those in which all detectors have the same detection efficiency) and zero background noise, the necessary and sufficient threshold detection efficiency for entangled qubits can be as low as $2/3$ for partially entangled states~\cite{Eberhard1993} and $0.828$ for maximally entangled states~\cite{Garg1987}. Massar~\cite{Massar2002} showed that high-dimensional systems could tolerate a detection efficiency that decreases with the dimension $d$ of the local quantum system. However, this result is of limited practical interest since an improvement over the qubit case occurs only for $d > 1600$. V\'ertesi, Pironio, and Brunner~\cite{Vertesi2010} identified a symmetric Bell inequality for which the efficiency can be lowered down to $0.618$ for partially entangled states and $0.77$ for maximally entangled states, using four-dimensional systems and assuming perfect visibility, which is still not sufficiently low for practical applications. Other proposals for loophole-free Bell tests with low detection efficiency either combine low-efficient detectors with nearly perfect ones~\cite{CL07,BGSS07,Garbarino10,AQCFCT12} or use more than five spatially separated parties~\cite{Larsson2001,CRV08,PVB12}, which is unpractical for real-world applications.

The critical detection efficiency $\eta$ is not the only important parameter in a loophole-free Bell experiment. Another essential variable is the required visibility $\vi$, which quantifies how much noise can be tolerated. 
The best combinations of parameters $(\eta,\vi)$ reported in photonic experiments in distances $\lessapprox 200$\,m are: $(0.774, 0.99)$~\cite{giustina2015significant}, $(0.763, 0.99)$~\cite{shalm2021device}, and $(0.8411, 0.9875)$~\cite{liu2021device}. However, these values are very difficult to achieve in longer distances.

In this work, we propose a general method to reduce the detection efficiency requirement \emph{exponentially} for any given Bell inequality. This is achieved by violating $N$ Bell inequalities in parallel with a source of $N$ entangled states carried by a single pair of particles. The value of the required detection efficiency then scales like $(\frac{C}{Q})^N$, where $C$ is the LHV bound, and $Q$ is a quantum value, i.e., the decay is exponential. Moreover, our method reduces the required detection efficiency for a given target visibility or a Bell inequality violation. We analyze in detail the case of parallel violation of $N$~Clauser-Horne-Shimony-Holt (CHSH) Bell inequalities~\cite{Clauser1969}. Another advantage of our approach is that the observed correlations can be directly used for practical applications, since the observed value of $N$~CHSH inequalities can be connected to the violation of an individual CHSH inequality. Hence, there is no need to develop new protocols based on Bell inequalities with more settings~\cite{Collins2004}.


{\em Physical setup.---}Consider a Bell experiment in which two spatially separated parties, Alice and Bob, have access to a source of high-dimensional entanglement carried by a single pair of particles. The key examples to keep in mind are hyperentangled states~\cite{Kwiat1997}, in which two photons are entangled across multiple degrees of freedom, and photon pairs entangled in high-dimensional degrees of freedom \cite{Erhard2020}.
Throughout the text, we consider photons as physical carriers of entanglement, however, similar reasoning can be applied to atoms, ions, etc.

Let us assume now that the carried high-dimensional entangled state is a product of $N$ entangled states, as it is the case for hyperentanglement~\cite{Kwiat1997}. We also assume that Alice and Bob can perform joint measurements on their subsystems producing $N$ outcomes each from a single click of their detectors. The main idea of the method is to use $N$ outcomes from each run of the experiment to violate $N$ Bell inequalities \emph{in parallel}. In this way, the probability of detectors' clicks for each of the $N$ inequalities is of the order of the $N$th root of the efficiency of the photon detection, i.e., it is effectively increased. We will provide a rigorous analysis that supports this claim. 

To the best of our knowledge, the conjecture that the critical detection efficiency could be lowered by integrating several qubit-qubit entangled states in one pair of particles was first made in Ref.~\cite{Barrett2002}, without a proof. In Ref.~\cite{Cabello2006}, it was shown that the critical detection efficiency could be reduced for the so-called Einstein-Podolsky-Rosen-Bell inequalities that require perfect correlations~\cite{Eberhard1995}. Similar ideas have been developed in later works focused on quantum key distribution~\cite{jain2020parallel,doda2021quantum} and the $p$-value of a Bell test~\cite{araujo2020bell}. Very recently, the idea has been explored for the case of 2-qubit maximally entangled states~\cite{marton2021bounding}. In this paper, we introduce a much more powerful and practical tool: penalized $N$-product (PNP) Bell inequalities. This tool leads to smaller critical detection efficiencies than those obtained in Ref.~\cite{marton2021bounding} and applies to any quantum violation of any Bell inequality, thus opening a new path toward loophole-free Bell tests with longer distances and higher dimensions.


{\em Product Bell inequalities.---}Let us consider $N$ Bell inequalities of the same type, evaluated ``in parallel''. Our first task is to identify a single parameter that quantifies the violation of local realism. One way to do it is to consider the product of the $N$ parameters of all $N$ Bell inequalities. Following this approach, let us start with a Bell inequality of the form 
\beq
\label{eq:bell}
\sum_{a,b,x,y}\Pr(a,b|x,y)c_{a,b}^{x,y} \leq C,
\eeq
with $c_{a,b}^{x,y}\geq 0$, 
where $\Pr(a,b|x,y)$ denotes the conditional probability of Alice and Bob to observe outcomes $a$ and $b$, respectively, given their choice of measurement settings $x$ and $y$, respectively, and $C$ is the LHV bound.
Our construction applies to \emph{any} Bell inequality, therefore, we will not specify the sets from which the settings or outcome take their values, we will just assume these sets to be finite.
Note, that any Bell inequality can be brought to the form of a ``nonlocal game'' in Eq.~\eqref{eq:bell} with $c_{a,b}^{x,y}\geq 0$~\cite{araujo2020bell}.

An $N$-\emph{product Bell inequality} based on Eq.~(\ref{eq:bell}) is defined as
\beq
\label{eq:bell_N}
\sum_{\va,\vb,\vx,\vy}\Pr(\va,\vb|\vx,\vy)\prod_{i=1}^N c_{a_i,b_i}^{x_i,y_i} \leq C_N,
\eeq
where $\va = (a_1,\dots,a_N)$ is a tuple of Alice's measurement outcomes, and $\vb,\vx,\vy$ similarly defined. $C_N$ denotes the maximum value of the $N$-product Bell inequality attainable by LHV models. 

One could expect that $C_N = C^N$. However, this is not the case for arbitrary Bell inequalities of the form given by Eq.~(\ref{eq:bell}), including the CHSH inequality~\cite{Clauser1969}. Indeed, for the CHSH inequality, $C=\frac{3}{4}$ but $C_2=\frac{10}{16}$~\cite{Barrett2002} and $C_3 = \frac{31}{64}$~\cite{araujo2020bell}.
This fact is also referred to as the \emph{simultaneous measurement loophole} in Bell tests~\cite{Barrett2002}. The problem of determining the closed form for $C_N$ (for the cases when $C_N>C^N$) is closely related to the so-called \emph{parallel repetition theorem} in interactive proof systems~\cite{Raz1998}. This problem was tackled in Ref.~\cite{Holenstein2007,Feige2007,Rao2011}, where only asymptotic upper-bounds on $C_N$ were reported. Moreover, the authors of Ref.~\cite{Feige2007} emphasized the difficulty of finding exact values of $C_N$. 

In this work, we take a different approach to the problem. Instead of trying to find the values of $C_N$, we propose a method for modifying the Bell expression in Eq.~(\ref{eq:bell_N}) in a way that $C_N = C^N$ holds for all $N$. We achieve this by adding a nonlinear ``penalty term'' to the left-hand side of Eq.~(\ref{eq:bell_N}), which forces a product local strategy (i.e., one in which each outcome $a_i$ depends only on $x_i$, and similarly for Bob) to be optimal. Given a Bell expression specified by coefficients $c^{x,y}_{a,b}\geq 0$ and the LHV bound $C$, we define a \emph{penalized $N$-product (PNP) Bell inequality} as follows:
\beq
\label{eq:bell_pen}
\sum_{\va,\vb,\vx,\vy}\Pr(\va,\vb|\vx,\vy)\prod_{i=1}^N c_{a_i,b_i}^{x_i,y_i} - \sum_{i=1}^{N-1}\kappa_iP_i \leq C^N,
\eeq
where $\kappa_i\in \mathbb{R}$ are some positive constants to be determined, and
\begin{equation}\label{eq:pen}
P_i = \sum_{\vx_i}\sum_{\vy_i}\sum_{a_i,b_i}\Big|\Pr(a_i,b_i|\vx_i,\vy_i)-\Pr(a_i,b_i|x_i,y_i)\Big|,
\end{equation}
where the summation is taken over all the values of the variables, and where we introduced an additional short-hand notation $\vx_i = (x_i,x_{i+1},\dots x_N)$, and similarly $\vy_i$, for $i\in \{1,\dots,N-1\}$.

The general idea of the method is rather straightforward. By taking large enough $\kappa_i$, we force the quantities $P_i$ for all $i\in \{1,\dots,N-1\}$ in Eq.~(\ref{eq:pen}) to be $0$. 
Once the penalty terms is zero, the LHV models are restricted to those in which the outcomes of Alice and Bob corresponding to $i$-th inequality are independent of the settings for inequalities $1$ to $i-1$.
This corresponds to a ``causal'' order for both paries in which the individual inequalities are evaluated, which is known to result in $C_N=C^N$. 

The remaining question is how large $\kappa_i$ need to be to ensure $C_N=C^N$, which we answer below.

\begin{result}
\label{th:bell_N} 
 Given a Bell inequality specified by coefficients $c^{x,y}_{a,b}\geq 0$, with $a,b,x$,$y$ taking values from some finite sets, it is sufficient to take $\kappa_i=\frac{1}{2}\Sigma C^{N-i}(\Sigma+C)^{i-1}$, $i\in \{1,\dots, N-1\}$, such that the LHV bound of the corresponding PNP Bell inequality is $C^N$, where $\Sigma = \sum_{a,b,x,y}c^{x,y}_{a,b}$ is the Bell's inequality algebraic bound.
\end{result}
The proof is given in Appendix~\ref{app:proof}.
The purpose of the upper bound on the sufficient values of $\kappa_i$ is not only theoretical. In practice, even if we use a product quantum strategy, due to experimental errors both $\mathrm{A}$ and $\mathrm{B}$ will have small yet nonzero values. These errors will be multiplied by $\kappa_i$ and could potentially result in no violation. 


{\em Lowering the critical detection efficiency.---}Here, we show that having a source of photon pairs carrying $N$ entangled states each alongside with PNP Bell inequalities allows for a significant reduction in the critical detection efficiency requirements for the violation of local realism.

To avoid the \emph{fair sampling assumption}~\cite{Branciard2011}, the parties need to either treat ``no-click" events as additional outcomes or employ a local assignment strategy~\cite{Czechlewski2018}. The latter means that whenever one party's detector does not click (when it should), the party draws an outcome according to some local (deterministic) strategy. This allows the parties to use the same Bell inequality without the need to find one with more outcomes.

In this work, we consider the local assignment strategy for mitigation of the ``no-click" events. 
Let $\bigotimes_{i=1}^N\rho_{AB}$ be a state carried by photon pair in out setup. Let $\bigotimes_{i=1}^N\mathcal{A}^{x_i}_{a_i}$ and $\bigotimes_{i=1}^N\mathcal{B}^{y_i}_{b_i}$ be the POVM (positive-operator valued measure) elements of Alice and Bob respectively, i.e., they are formed by the POVM elements $\mathcal{A}^{x}_{a}$ and $\mathcal{B}^{y}_{b}$, that are the same for all $i$. Evidently, this leads to quantum behavior of the form $\Pr(\va,\vb|\vx,\vy) = \prod_{i=1}^N\tr(\mathcal{A}^{x_i}_{a_i}\otimes\mathcal{B}^{y_i}_{b_i}\rho_{AB})$. Let $\alpha$ and $\beta$ be deterministic assignment strategies $a_i = \alpha(x_i)$ and $b_i = \beta(y_i)$, for all $i$, employed by Alice and Bob respectively in case of a ``no-click" event. If, for instance, Bob's detector does not click but Alice's does, the parties observed behavior is $\Pr(\va,\vb|\vx,\vy) = \prod_{i=1}^N\tr(\mathcal{A}^{x_i}_{a_i}\rho_A)\delta_{b_i,\beta(y_i)}$, where $\rho_A$ is Alice's reduced state $\rho_{AB}$ and $\delta_{\cdot,\cdot}$ is the Kronecker delta. Similarly, the parities observe the behavior $\Pr(\va,\vb|\vx,\vy) = \prod_{i=1}^N\tr(\mathcal{B}^{y_i}_{b_i}\rho_B)\delta_{a_i,\alpha(x_i)}$ whenever Alice's detector does not click, but the one of Bob does. Finally, for the cases of no clicks on both detectors, the parties observe a local deterministic behavior $\Pr(\va,\vb|\vx,\vy) = \prod_{i=1}^N\delta_{a_i,\alpha(x_i)}\delta_{b_i,\beta(y_i)}$.


\begin{figure}[t!]\centering
 \includegraphics[width=0.48\textwidth]{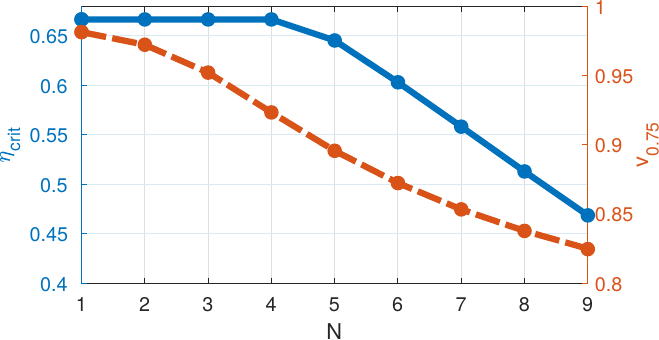}
 \caption{(Solid line) Critical detection efficiency $\eta_{\text{crit}}$ for the PNP Bell inequality as a function of $N$. (Dashed line) Visibility (per qubit pair) required for a loophole-free Bell test when $\eta=0.75$ as a function of $N$. Perfect statistics is assumed, i.e., the penalty term is assumed to be zero.}\label{fig:eta_crit}
\end{figure}


Assuming the detection efficiency of Alice's and Bob's detectors to be $\eta$, the value of the PNP Bell expression is the following:
\beq
\label{eq:bell_eta}
\eta^2Q^N+\eta(1-\eta)(A^N+B^N)+(1-\eta)^2C^N,
\eeq
with
\begin{equation}
\label{eq:local_rand}
\begin{split}
Q &= \sum_{a,b,x,y}c_{a,b}^{x,y}\tr(\mathcal{A}^x_a\otimes\mathcal{B}^y_b\rho_{AB}),\\
A &= \sum_{a,b,x,y}c_{a,b}^{x,y}\tr(\mathcal{A}^x_a\rho_{A})\delta_{b,\beta(y)},\\
B &= \sum_{a,b,x,y}c_{a,b}^{x,y}\tr(\mathcal{B}^y_b\rho_{B})\delta_{a,\alpha(x)},
\end{split}
\end{equation}
where we have assumed that the local strategies $\alpha$ and $\beta$ reproduce the LHV bound $C$. Clearly, since all the aforementioned strategies are product, the penalty term is exactly $0$. Notice that in Eq.~(\ref{eq:bell_eta}), $\eta$ appears only in its second power, precisely due to the fact that the $N$-qudit state $\bigotimes_{i=1}^N\rho_{AB}$ is carried by a single pair of photons. This is what we meant when we said that the effective detection efficiency for each of the $N$ Bell inequalities is of the order of $\eta^{\frac{1}{N}}$.

To observe a violation of local realism, one needs to ensure that the value of the expression in Eq.~(\ref{eq:bell_eta}) is greater than the LHV bound $C^N$. Solving this inequality with respect to $\eta$, we obtain the following value of the required detection efficiency for given $Q$, $A$, and $B$:
\beq
\label{eq:eta_crit}
\eta = \frac{2C^N-A^N-B^N}{Q^N+C^N-A^N-B^N}.
\eeq
This equation has the following interesting implication.
\begin{remark}
For any given Bell inequality with binary outcomes and a quantum strategy with $Q>C$, it follows from Eq.~(\ref{eq:eta_crit}) that the detection efficiency requirement decays exponentially with $N$.
\end{remark}
\noindent Indeed, if we take $A = B = \delta C$, then $\eta = 2\left(\frac{C}{Q}\right)^N(1-\delta^N)+O\left(\left(\frac{C}{Q}\right)^{2N}\right)$. For any Bell inequality, $\delta<1$ whenever $Q>C$. Hence, the decay of $\eta$ with $N\to\infty$ is at least exponential with the factor of $\log(\frac{C}{Q})$. The above remark is in parallel with the results of Massar~\cite{Massar2002}. It is worth noting that the critical detection efficiency in the Massar construction decreases exponentially with dimension, but necessitates an exponentially large number of measurement settings. Our method provides a polynomial in dimension, i.e., an exponential in the number of qubits, reduction in critical detection efficiency, but it only requires a polynomial number of measurement settings.

In order to find the critical detection efficiency $\eta_{\text{crit}}$ for a given Bell inequality and its corresponding PNP Bell inequality, one needs to optimize $\eta$ in Eq.~(\ref{eq:eta_crit}) over all possible values of $(Q,A,B)$. In what follows, we solve this optimization problem for the $N$-product CHSH inequality. 


\begin{figure}[t!]\centering
 \includegraphics[width=0.46\textwidth]{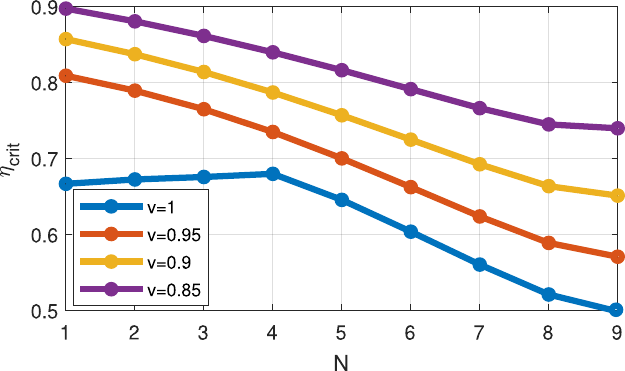}
 \caption{Required detection efficiency for different values of the visibility $\vi$, as a function of $N$. Lines in the plot are arranged from bottom to top as $\vi$ changes from $1$ to $0.85$. To see the effect of the penalty term, each difference between the probabilities in Eq.~(\ref{eq:pen}) is taken to be $10^{-7}$, and $\kappa_i$ are chosen according to Result 1.}\label{fig:eta_real}
\end{figure}


{\em PNP inequality for the CHSH inequality.---}The coefficients of the CHSH inequality~\cite{Clauser1969} in its nonlocal game formulation are $c^{x,y}_{a,b} = \frac{1}{4}\delta_{a\oplus b,xy}$, where $a,b,x,y\in \{0,1\}$ and $\oplus$ denotes addition modulo $2$. For this form of the CHSH inequality, we have $C=\frac{3}{4}$, the quantum bound $Q_{\text{max}} = \frac{1}{2}+\frac{1}{2\sqrt{2}}$, and $\Sigma = 1$. In order to minimize the expression in Eq.~(\ref{eq:eta_crit}) over all quantum states and measurements, first we determine the maximal values of $A$ and $B$ attainable for a given value of $Q$, and then optimize Eq.~(\ref{eq:eta_crit}) over $Q$. In particular, due to the symmetry with respect to $A$ and $B$ in Eq.~(\ref{eq:eta_crit}), we are interested in the situation $A=B$. For this case, the optimal relation is the following:
\beq
\label{eq:opt_ab}
A = B = \frac{1}{2}+\frac{1}{4}\sqrt{(1-q)\left(1+\frac{q}{\sqrt{1+q^2}}\right)},
\eeq
where $q = \sqrt{(4Q-2)^2-1}$. As $Q$ changes from $\frac{3}{4}$ to $\frac{1}{2}+\frac{1}{2\sqrt{2}}$, $q$ increases from $0$ to $1$, and, hence, $A$ and $B$ decrease from $\frac{3}{4}$ to $\frac{1}{2}$. For the $2$-qubit state $\rho_{AB}$ and qubit measurements $\mathcal{A}^x_a$ and $\mathcal{B}^y_b$ that produce the relation in Eq.~(\ref{eq:opt_ab}) see Appendix~\ref{app:chsh}. We used the Navascu\'es-Pironio-Ac\'{\i}n hierarchy~\cite{Navascues2007} to indicate the dimension-independent optimality of Eq.~(\ref{eq:opt_ab}).

Employing the relation in Eq.~(\ref{eq:opt_ab}), we optimize $\eta$ in Eq.~(\ref{eq:eta_crit}) over $Q$ in order to obtain the optimal value $\eta_{\text{crit}}$ for a given $N$. We plot the results in Fig.~\ref{fig:eta_crit}. In the same figure, we show the minimal visibility $\vi_{0.75}$ for which violation can still be observed with detectors of a given detection efficiency $\eta=0.75$. As we can see, even though taking $2$, $3$, and $4$-product CHSH inequalities does not decrease the value of $\eta_{\text{crit}}$, one can obtain a significant advantage in terms of visibility for $\eta>\eta_{\text{crit}}$. 

In Fig.~\ref{fig:eta_real} we plot $\eta_\vi$, the required detection efficiency to observe a violation of the PNP Bell inequality with visibility as low as $\vi$. We also account for possible experimental imperfections by taking nonzero values of the penalty term. 


{\em Summary and outlook.---}In this work, we addressed the problem of reducing the detection efficiency requirements for loophole-free Bell experiments in order to achieve loophole-free Bell tests over longer distances. We presented a method that, when applied to any given Bell inequality, produces a new Bell inequality by taking a penalized product of $N$~copies of it, for which the critical detection efficiency decays exponentially with~$N$. This implies that the critical detection efficiency can be drastically reduced in experiments using photon sources that allow for encoding multiple copies of a qubit-qubit (or qudit-qudit) entangled state on a single pair of particles. Examples of such sources are hyperentanglement sources and sources of high-dimensional entanglement.

We applied our method to several binary Bell inequalities and found that the lowest detection efficiencies occur for the PNP CHSH inequality. The advantage of the CHSH inequality is in terms of both critical detection efficiency and visibility of the violation. A natural target for future work is to identify Bell inequalities for which the critical detection efficiencies are low enough for mid-distance photonic loophole-free Bell tests and related applications such as device-independent quantum key distribution.
Another relevant problem is the calculation of $p$-values for PNP Bell inequalities, which would depend on the value of the penalty term. Finally, it is interesting to see whether PNP Bell inequalities can be used for a single-shot Bell test~\cite{araujo2020bell}.

{\em Note added.---} A few years after completion of this work, some authors of this article have identified that the identified relation in Eq.~\eqref{eq:opt_ab} is sub-optimal, and a slightly better strategy exists in quantum mechanics, which, although does not seem to admit a concise algebraic formulation. 


\begin{acknowledgements}
We would like to thank Will McCutcheon, Konrad Banaszek, Costantino Budroni, Mateus Ara\'u{}jo, Jean-Marc Merolla, Miguel Navascu\'{e}s, and Marek \.Zukowski, for fruitful discussions and comments.
This research was made possible by funding from QuantERA, an ERA-Net cofund in Quantum Technologies (www.quantera.eu), under projects SECRET (MINECO Project No.~PCI2019-111885-2) and eDICT. We also acknowledge the financial support by First TEAM Grant No.~2016-1/5 and Project Qdisc (Project No.~US-15097), with FEDER funds. This research was funded by the Deutsche Forschungsgemeinschaft (DFG, German Research Foundation), Project No.~441423094. M.\,P.~acknowledges support by the Foundation for Polish Science (IRAP project, ICTQT, Contract No.~2018/MAB/5, cofinanced by EU within Smart Growth Operational Programme). A.\,Ch.~acknowledges support by the NCN grant SHENG (Contract No.~2018/30/Q/ST2/00625). M.\,B.~acknowledge support by the Knut and Alice Wallenberg Foundation and  the Swedish  Research  Council. A.\,C. acknowledges
support from the Knut and Alice Wallenberg Foundation through the Wallenberg Centre for Quantum Technology
(WACQT). The numerical optimization was carried out using \href{https://ncpol2sdpa.readthedocs.io/en/stable/index.html}{Ncpol2sdpa} \cite{wittek2015algorithm}, \href{https://yalmip.github.io/}{YALMIP} \cite{Lofberg2004}, \href{https://www.mosek.com/documentation/}{MOSEK} \cite{mosek}, and \href{https://cvxopt.org/}{CVXOPT} \cite{andersen2013cvxopt}.
\end{acknowledgements}
 

\newpage




\onecolumngrid
\begin{appendix}
\section*{Appendix}

\section{Proof of Result 1}
\label{app:proof}
\begin{proof}
    Let us start by writing the expression of the PNP Bell inequality in Eq.~\eqref{eq:bell_pen},
    \begin{equation}\label{eq:bell_pen_app}
        \sum_{\va,\vb,\vx,\vy}\Pr(\va,\vb|\vx,\vy)\prod_{i=1}^N c_{a_i,b_i}^{x_i,y_i} - \sum_{i=1}^{N-1}\kappa_i\sum_{\vx_i}\sum_{\vy_i}\sum_{a_i,b_i}\Big|\Pr(a_i,b_i|\vx_i,\vy_i)-\Pr(a_i,b_i|x_i,y_i)\Big|,
    \end{equation}
and let $C$ be the LHV bound for a single copy of the original Bell inequality, and $\Sigma = \sum_{a,b,x,y}c_{a,b}^{x,y}$ be the algebraic bound.

Let us define for every $i\in \{1,\dots N-1\}$,
\begin{equation}\label{eq:veps_i}
    \veps_i = \max_{a_i,b_i,\vx_{i},\vy_{i}}\left\{\Big|\Pr(a_i,b_i|\vx_i,\vy_i)-\Pr(a_i,b_i|x_i,y_i)\Big|\right\},
\end{equation}
which quantifies the maximal deviation of the distribution from the product strategy for the $i$-th copy.
We can upper-bound the first part of the expression in Eq.~\eqref{eq:bell_pen_app} as
\begin{equation}
    \begin{split}
        \sum_{\va,\vb,\vx,\vy}\Pr(\va,\vb|\vx,\vy)\prod_{i=1}^N c_{a_i,b_i}^{x_i,y_i} & = \sum_{a_1,b_1,x_1,y_1}c_{a_1,b_1}^{x_1,y_1}\sum_{\vx_2,\vy_2}\Pr(a_1,b_1|\vx,\vy)\sum_{\va_2,\vb_2}\Pr(\va_2,\vb_2|\vx,\vy,a_1,b_1)\prod_{i=2}^N c_{a_i,b_i}^{x_i,y_i}\\
        & \leq  \sum_{a_1,b_1,x_1,y_1}c_{a_1,b_1}^{x_1,y_1}\Big(\veps_1+\Pr(a_1,b_1|x_1,y_1)\Big)\sum_{\va_2,\vb_2,\vx_2,\vy_2}\Pr(\va_2,\vb_2|\vx,\vy,a_1,b_1)\prod_{i=2}^N c_{a_i,b_i}^{x_i,y_i}\\
        & \leq \left(\veps_1\Sigma+C\right)\max_{a_1,b_1,x_1,y_1}\left\{\sum_{\va_2,\vb_2,\vx_2,\vy_2}\Pr(\va_2,\vb_2|\vx_2,\vy_2,a_1,b_1,x_1,y_1)\prod_{i=2}^N c_{a_i,b_i}^{x_i,y_i}\right\}.
    \end{split}
\end{equation}
Next, we use the fact that conditioning on the particular values of $a_1,b_1,x_1$, and $y_1$ does not change the bound of the remaining $N-1$ copies of the Bell inequality.
Therefore, we can apply the same reasoning as above for the copies $i\in\{2,\dots,N-1\}$, and obtain the following upper-bound
\begin{equation}
    \sum_{\va,\vb,\vx,\vy}\Pr(\va,\vb|\vx,\vy)\prod_{i=1}^N c_{a_i,b_i}^{x_i,y_i} \leq \prod_{i=1}^{N-1}\left(\veps_i\Sigma+C\right)C\ ,
\end{equation}
where the last copy of the Bell inequality can be upper-bounded simply by $C$.
The penalty term in Eq.~\eqref{eq:bell_pen_app} can be lower-bounded by $2\sum_{i=1}^{N-1}\kappa_i\veps_i$, where the factor of $2$ is due to the normalization of the probability distributions.
Having bounded both parts of the expression  in Eq.~\eqref{eq:bell_pen_app} in terms of $\veps_i$, we can determine the values of $\kappa_i$ for which the LHV bound for the penalized Bell inequality is $C^N$, by requiring that
\begin{equation}
    \prod_{i=1}^{N-1}\left(\veps_i\Sigma+C\right)C-2\sum_{i=1}^{N-1}\kappa_i\veps_i\leq C^N\ .
\end{equation}
Indeed, if we take $\kappa_i=\frac{1}{2}\Sigma C^{N-i}(\Sigma+C)^{i-1}$ for $i\in\{1,\dots,N-1\}$, as in the statement of the results, we can iteratively show that
\begin{equation}
    \begin{split}
    \prod_{i=1}^{N-1}\left(\veps_i\Sigma+C\right)C-2\sum_{i=1}^{N-1}\kappa_i\veps_i & = \prod_{i=1}^{N-2}(\veps_i\Sigma+C)C^2-2\sum_{i=1}^{N-2}\kappa_i\veps_i+\veps_{N-1}\left(\Sigma C  \prod_{i=1}^{N-2}(\veps_i\Sigma+C)-2\kappa_{N-1}\right)\\
        &\leq \prod_{i=1}^{N-3}(\veps_i\Sigma+C)C^3-2\sum_{i=1}^{N-3}\kappa_i\veps_i+\veps_{N-2}\left(\Sigma C^2\prod_{i=1}^{N-3}(\veps_i\Sigma+C)-2\kappa_{N-2}\right)\\
        & \leq \dots \leq (\veps_1\Sigma+C)C^{N-1}-2\kappa_1\veps_1\leq C^N\ .
    \end{split} 
\end{equation}
\end{proof}

\section{Optimal states and measurements for the CHSH inequality}
\label{app:chsh}


Here, we give exact form of the optimal states and measurements that reproduce the relation form Eq.~(\ref{eq:opt_ab}) between $A$, $B$, and $Q$ [Eqs.~(\ref{eq:local_rand})] for the CHSH inequality. 
The state $\rho_{AB}$ is the following nonmaximally entangled pure state:
\begin{equation}
\rho_{AB} = \frac{1+\sqrt{1-q^2}}{2}\ketbra{00}{00}+\frac{1-\sqrt{1-q^2}}{2}\ketbra{11}{11}+\frac{q}{2}\Big(\ketbra{00}{11}+\ketbra{11}{00}\Big).
\end{equation}
The observables $\mathcal{A}^x_0-\mathcal{A}^x_1$ and $\mathcal{B}^y_0-\mathcal{B}^y_1$ are sharp and given by
\begin{equation}\begin{split}
\mathcal{A}^x_0-\mathcal{A}^x_1 &= \sqrt{\frac{1}{1+q}\left[1+\frac{(-1)^xq}{\sqrt{1+q^2}}\right]}\sigma_z+(-1)^x\sqrt{\frac{q}{1+q}\left[1-\frac{(-1)^x}{\sqrt{1+q^2}}\right]}\sigma_x, \quad x\in\{0,1\},\\
\mathcal{B}^y_0-\mathcal{B}^y_1 &= \sqrt{\frac{1}{1+q}\left[1+\frac{(-1)^yq}{\sqrt{1+q^2}}\right]}\sigma_z-(-1)^y\sqrt{\frac{q}{1+q}\left[1-\frac{(-1)^y}{\sqrt{1+q^2}}\right]}\sigma_x, \quad y\in\{0,1\},
\end{split}\end{equation}
where $\sigma_z$ and $\sigma_x$ are the Pauli $z$ and $x$ matrices, respectively. The parameter $q$ is the one in Eq.~(\ref{eq:opt_ab}) that gives the values of $A$ and $B$. The value $Q$ of the CHSH inequality is $Q = \frac{1}{2}+\frac{1}{4}\sqrt{1+q^2}$.
\end{appendix}


\twocolumngrid
\bibliographystyle{apsrev4-1}
\bibliography{par_bell}
 
\end{document}